\documentclass[review, sort&compress, hidelinks]{elsarticle}

\usepackage{textcomp}
\usepackage{lineno}
\usepackage[hidelinks]{hyperref}
\usepackage[utf8]{inputenc}
\usepackage[english]{babel}
\modulolinenumbers[5]
\usepackage{floatrow}
\usepackage{physics}
\usepackage{graphicx}
\usepackage{amssymb}
\usepackage{amsmath}
\usepackage{xcolor}
\usepackage{comment}
\usepackage{url}
\usepackage{enumerate}
\usepackage{caption}
\usepackage{subcaption}
\usepackage{amssymb}
\usepackage{mathrsfs}
\usepackage{booktabs}
\usepackage{multirow}
\usepackage[margin=2.5cm]{geometry}
\usepackage{array,longtable}
\usepackage{makecell}
\usepackage{tabu, booktabs}
\usepackage{mathtools}

\usepackage{adjustbox}
\usepackage{algorithm}
\usepackage{algpseudocode}

\usepackage{pgfplots}
\usetikzlibrary{arrows}
\usepackage{booktabs}
\usepackage{multirow}
\usepackage{xcolor}
\usepackage{color}
\usepackage{microtype}
\usepackage{tabularx}
\usepackage[numbers]{natbib}
\usepackage{upgreek} %Gc

\usepackage{shellesc}
\usepackage{tikz}
\usetikzlibrary{calc}
\usepackage[T1]{fontenc}
\usepackage{lmodern}

\usepackage{siunitx}
\usepackage{mathdots}
\usepackage{yhmath}
\usepackage{gensymb}
\usepackage{wasysym}

\usetikzlibrary{fadings}
\usetikzlibrary{patterns}
\usetikzlibrary{shadows.blur}
\usetikzlibrary{arrows}
\usetikzlibrary{arrows.meta}

\makeatletter
\def\ps@pprintTitle{%
	\let\@oddhead\@empty
	\let\@evenhead\@empty
	\def\@oddfoot{\reset@font\hfil\thepage\hfil}
	\let\@evenfoot\@oddfoot
}
\makeatother

\begin{document}

\begin{frontmatter}

\title{Cyclic viscoelastic-viscoplastic behavior of epoxy nanocomposites under hygrothermal conditions: A phase-field fracture model}

\author[firstaddress]{Behrouz Arash\corref{mycorrespondingauthor}}
\ead{behrouza@oslomet.no}
\author[firstaddress]{Shadab Zakavati}
%\ead{b.arash@isd.uni-hannover.de}
\author[secondaddress]{Betim Bahtiri}
%\ead{b.bahtiri@isd.uni-hannover.de}
\author[thirdaddress]{Maximilian Jux}
\author[secondaddress]{Raimund Rolfes}
%\ead{r.rolfes@isd.uni-hannover.de}

\address[firstaddress]{Department of Mechanical, Electrical, and Chemical Engineering, Oslo Metropolitan University, 0166 Oslo, Norway}
\address[secondaddress]{Institute of Structural Analysis, Leibniz Universit{\"a}t Hannover, Appelstra{\ss}e 9A, 30167 Hannover, Germany}
\address[thirdaddress]{Institute of Composite Structures and Adaptive Systems, DLR (German Aerospace Center), Lilienthalplatz 7, 38108 Brunswick, Germany}

%------------------------------------------------------------------------------

\begin{abstract}
In this study, a finite deformation phase-field formulation is developed to investigate the effect of hygrothermal conditions on the viscoelastic-viscoplastic fracture behavior of epoxy nanocomposites under cyclic loading. The formulation incorporates a definition of the Helmholtz free energy, which considers the effect of nanoparticles, moisture content, and temperature. The free energy is additively decomposed into a deviatoric equilibrium, a deviatoric non-equilibrium, and a volumetric contribution, with distinct definitions for tension and compression. The proposed derivation offers a realistic modeling of damage and viscoplasticity mechanisms in the nanocomposites by coupling the phase-field damage model with a modified crack driving force and a viscoelastic-viscoplastic model. Numerical simulations are conducted to study the cyclic force-displacement response of both dry and saturated boehmite nanoparticle (BNP)/epoxy samples, considering BNP contents and temperature. Comparing numerical results with experimental data shows good agreement at various BNP contents. In addition, the predictive capability of the phase-field model is evaluated through simulations of single-edge notched nanocomposite plates subjected to monolithic tensile and shear loading.
\end{abstract}


\end{frontmatter}

%\linenumbers

%------------------------------------------------------------------------------

\section{Introduction}
\label{intro}

In the field of engineering, a major challenge is reducing the weight of structures to improve their performance and functionality for specific applications. To achieve this, researchers are focusing on both optimizing the structure and developing new materials that have superior thermo-mechanical properties but are lightweight. One such material is polymer nanocomposites, which combine the desirable attributes of polymers, such as low weight and high ductility, with the unique features of nanoparticles~\cite{spitalsky2010carbon,zhou2019interface,li2016study,mousavi2021optimization}. Studies have shown that boehmite nanoparticle (BNP) reinforced epoxy composites are among the most promising composites for lightweight structures due to their high strength-to-weight ratio~\cite{jux2018mechanical}. Compared to neat epoxies, BNP/epoxy nanocomposites have significantly improved mechanical properties, including strength and fracture toughness~\cite{jux2018mechanical,khorasani2019effect,arash2019viscoelastic}.

To advance the material innovation, reliable models are required to predict how external conditions (such as loading rate, temperature, and moisture) and microstructural parameters (such as nanoparticle/matrix interactions) affect the material behavior of nanocomposites. Continuing research activity on polymers and their composites has led to a variety of phenomenological or physically motivated constitutive models~\cite{bardella2001phenomenological,zhou2007experimental,vogler2013modeling,vu2015multiscale,nguyen2016large,park2018toward} to elucidate their nonlinear rate- and temperature-dependent behavior. Boyce et al.~\cite{boyce2000constitutive} developed a constitutive model based on a composite-type formulation considering the microstructure of semicrystalline polymers. In the model, the soft amorphous and stiff crystalline phases are treated as the matrix and fillers, respectively. Later, based on the model, Qi and Boyce~\cite{qi2005stress} proposed a viscoelastic-viscoplastic constitutive model to capture the nonlinear, rate-dependent, and softening behavior of thermoplastic polyurethanes. Li et al.~\cite{li2016molecular} introduced a physically-based viscoelastic constitutive model for elastomers at large deformation, where elastomers are assumed to be cross-linked networks with superimposed free chains. Nguyen et al.~\cite{nguyen2016large} developed and experimentally calibrated a rate-dependent damage constitutive model for epoxy resins to study the nonlinear behavior of amorphous glassy polymers. Based on the definition of Helmholtz free energy, N’Guyen et al.~\cite{n2016thermodynamical} derived a thermodynamical framework for the thermo-chemo-mechanical couplings in polymer materials at finite deformation.

However, predicting the nonlinear stress-strain response of polymer nanocomposites is more challenging due to the heterogeneous distribution of agglomerated nanoparticles in the matrix and complex interactions between the matrix and nanoparticles. To overcome these challenges, Fankh{\"a}nel et al.~\cite{fankhanel2019elastic} developed an atomistically-informed finite element (FE) model to investigate the material properties of BNP/epoxy nanocomposites. They used molecular simulations to characterize the interphase properties between BNPs and epoxy matrices, which were then scaled up to the continuum level. FE simulations of representative volume elements of the nanocomposites were then performed to homogenize the effective material properties. Arash et al.~\cite{arash2019viscoelastic,arash2019viscoelastic2} proposed a multiscale framework to calibrate a viscoelastic damage model for BNP/epoxy nanocomposites at finite deformation. The resulting model was validated through experimental-numerical testing, demonstrating its ability to accurately capture the stress-strain relationship of the nanocomposites, including nonlinear hyperelastic, rate-dependent, and softening behavior. Unger et al.~\cite{unger2019non,unger2020effect} extended the framework to characterize the thermo-viscoelastic damage behavior of BNP/epoxy nanocomposites and developed a robust parameter identification procedure. 

When it comes to computational modeling of fracture in polymer nanocomposites, accurately predicting crack initiation and propagation is crucial. Research has shown that cracks are related to the generation of microvoids and microcracks in loaded polymer materials~\cite{chowdhury2008effects,nguyen2016large}. As loading continues, these microvoids and microcracks will coalesce, leading to the birth of complete cracks. %This causes deformation to localize into a narrow zone, accompanied by a softening behavior. 
The damage mechanisms are also affected by nanoparticle contents and hygrothermal conditions~\cite{bahtiri2022elucidating,BAHTIRI2023116293,arash2023effect}. However, FE modeling of the damage mechanisms based on local continuum description of damage typically suffer from an inherent mesh dependence~\cite{needleman1988material,geers1998strain}. 

To overcome this issue and investigate damage and failure in materials, regularized solutions have been proposed in the literature. Various models have been developed based on regularization theories, such as the gradient-enhanced damage model and its variations~\cite{peerlings1996gradient,saroukhani2013simplified,poh2017localizing,vandoren2018modeling,arash2021finite}. These models include the phase-field model (PFM)~\cite{francfort1998revisiting,aranson2000continuum,miehe2010thermodynamically}, which provides variational fracture models by minimizing potential energy consisting of the bulk energy, external forces work, and surface energy. PFMs use a scalar phase-field parameter to describe the smooth transition from an intact material to a fully broken state, making them an alternative to discontinuous crack modeling. They are capable of predicting complex patterns of crack initiation, propagation, and branching. PFMs have been used to study brittle, quasi-brittle, and ductile fracture~\cite{miehe2010phase,bourdin2000numerical,fang2020phase,wu2017unified,ambati2016phase,miehe2016phase,kumar2022nonlinear,dammass2023phase}. Shanthraj et al.~\cite{shanthraj2016phase} formulated a PFM for elasto-viscoplastic materials, which gives a physically realistic description of the material behaviour at the crack tip. Dean at al.~\cite{dean2020multi} developed a PFM for long fiber polymer composites by encompassing the differentiation of fiber and matrix failure phenomena. The formulation incorporates plastic effects via an invariant-based plasticity model for matrix-dominated deformation states. Msekh et al.~\cite{msekh2016predictions} developed a PFM to predict the tensile strength and fracture toughness of clay/epoxy nanocomposites, while Goswami et al.~\cite{goswami2020transfer,goswami2020adaptive} proposed a neural network algorithm for phase-field modeling of fracture in brittle materials. The simulation results show that the proposed approach can match the crack paths reported in the literature.

Furthermore, some phase-field formulations have been derived to study the rate-dependent fracture of solids~\cite{shen2019fracture,loew2019rate,yin2020fracture,brighenti2021phase}. In the case of polymer nanocomposites, Arash et al.~\cite{arash2021finite,arash2023effect} developed a phase-field formulation to study the nonlinear viscoelastic fracture behavior of BNP/epoxy nanocomposites at finite deformation, considering the effect of hygrothermal conditions. The formulation takes into account the effect of temperature, moisture, and nanoparticle contents, and defines the Helmholtz free energy through an additive decomposition of energy into equilibrium, non-equilibrium, and volumetric contributions, with varying definitions under compressive and tensile loading.

In addition to the above-mentioned inelastic mechanisms (i.e., nonlinear viscoelasticity and damage), experimental observations suggest the presence of additional irreversible deformation in nanocomposites subjected to cyclic load-unloading~\cite{BAHTIRI2023116293}. This indicates that the nanocomposites undergo viscoplastic deformation, which could significantly affect their overall behavior. Hence, for the more realistic modeling of these materials, a phase-field formulation, which incorporates both nonlinear viscoelasticity and viscoplasticity, is necessary.

To address the issue, this study focuses on developing a PFM to investigate the cyclic viscoelastic-viscoplastic fracture behavior of BNP/epoxy nanocomposites under hygrothermal conditions. For this, a PFM with a modified crack driving force is developed, considering the effect of nanoparticles, moisture content, and temperature. Meanwhile, dogebone tensile tests under dry and saturated conditions are conducted to calibrate and validate the proposed PFM at various BNP contents. The numerical-experimental comparison reveals that the PFM can accurately predict the damage and viscoplasticity mechanisms in the nanocomposites. Furthermore, the predictive capability of the model is qualitatively evaluated through simulations of single-edge notched nanocomposite plates subjected to monolithic tensile and shear loading.

This work is organized as follows. Sect.~\ref{sec:const} presents a viscoelastic-viscoplastic model describing the rate- and temperature-dependent behavior of the BNP/epoxy nanocomposites at finite deformation. In Sect.~\ref{sec:phase_model}, the governing equations of the PFM and the corresponding weak form and discretized equations are provided. In Sect.~\ref{sec:results}, numerical simulations are presented to validate the proposed PFM and study the effect of nanoparticle content, moisture and temperature on the fracture evolution in the nanocomposites. Finally, sect.~\ref{sec:summary} summarizes the findings.

%------------------------------------------------------------------------------

\section{Constitutive model for nanoparticle/epoxy}
\label{sec:const}

In this section, a viscoelastic-viscoplastic model for BNP/epoxy nanocomposites is presented. The stress response comprises of an equilibrium, two viscous, and a volumetric component to capture the nonlinear rate-dependent behavior of materials. The effect of BNPs and moisture on the stress-strain relationship is taken into account by defining an amplification factor as a function of the nanofiller and moisture contents. A modified Kitagawa model is also adopted to account for the effect of temperature. %Here, we also take into account the material swelling through moisture. 

\subsection{Kinematics}
\label{subsec:kinematics}

The total deformation gradient, containing the mechanical deformation, is multiplicatively split into a volumetric and deviatoric part as
\begin{align}
	\mathbf{F} & =\ J^{1/3} \bar{\mathbf{F}},
\end{align}
where $J = \text{det}[\mathbf{F}] $ and $\bar{\mathbf{F}}$ are the volumetric deformation and the deviatoric deformation gradient, respectively. The volume deformation is further decomposed into two terms: The mechanical compressibility $J_{m}$ and the moisture-induced swelling $J_{w}$, leading to an overall volumetric deformation as
\begin{align}
	J & = J_{m} J_{w},
\end{align}
where
\begin{align}\label{eq:Jw}
	J_{w} & = 1 + \alpha_{w} w_{w}.
\end{align}

In the equation above, $\alpha_{w}$ is the moisture swelling coefficient and $w_{w}$ is the moisture content~\cite{arash2023effect}. Our model incorporates experimental characteristics by decomposing the material behavior into a viscoelastic and a viscoplastic part, corresponding to the time-dependent reversible and time-dependent irreversible response, respectively. We further decompose the viscoelastic stress response into a hyperelastic network and a viscous network. The hyperelastic spring, associated with the entropy change due to deformations, captures the equilibrium response, while the viscous network composed of an elastic spring and a viscoelastic dashpot describes the non-equilibrium behavior of the nanocomposites. %Additionally, the quasi-irreversible sliding of the molecular chains, resulting in stress softening, also known as the Mullins effect~\cite{mullins1965stress,mullins1969softening}, is implemented within the constitutive model. 

The deviatoric part of the deformation gradient is decomposed into a viscoplastic and a viscoelastic component~\cite{govindjee1997presentation}:
\begin{align}\label{eq: dec_f}
	\bar{\mathbf{F}} & = \bar{\mathbf{F}}_{ve} . \bar{\mathbf{F}}_{vp}.
\end{align}

Also, the viscoelastic deformation gradient is split into an elastic and an inelastic part as
\begin{align}\label{eq: dec_ve}
	\bar{\mathbf{F}}_{ve} & = \bar{\mathbf{F}}_{e} . \bar{\mathbf{F}}_{v}.
\end{align}

Accordingly, similar decompositions are obtained for the left Cauchy-Green deformation tensors:
\begin{align}
	\bar{\mathbf{B}} & = \bar{\mathbf{F}} . \bar{\mathbf{F}}^{T},\\
	\bar{\mathbf{B}}_{e} & = \bar{\mathbf{F}}_{e} . \bar{\mathbf{F}}_{e}^{T},\\
	\bar{\mathbf{B}}_{ve} & = \bar{\mathbf{F}}_{ve} . \bar{\mathbf{F}}_{ve}^{T}.
\end{align}

The total velocity gradient of the viscoelastic network, $\bar{\mathbf{L}}_{ve} = \dot{\bar{\mathbf{F}}}_{ve}\left(\bar{ \mathbf{F}}_{ve}\right)^{-1}$, can be decomposed into an elastic and a viscous component analogously to Eq.~\eqref{eq: dec_ve}
\begin{align}
	\bar{\mathbf{L}}_{ve} & = \bar{\mathbf{L}}_{e} + \bar{\mathbf{F}}_{e}.\bar{\mathbf{L}}_{e}.\bar{\mathbf{F}}_{e}^{-1} = \bar{\mathbf{L}}_{e} + \tilde{\mathbf{L}}_{v},
\end{align}
and
\begin{align}\label{eq: l_v}
	\tilde{\mathbf{L}}_{v} & = \dot{\bar{\mathbf{F}}}_{v} . \bar{\mathbf{F}}_{v}^{-1} = \tilde{\mathbf{D}}_{v} + \tilde{\mathbf{W}}_{v}.
\end{align}

Here, $\tilde{\mathbf{D}}_{v}$ represents the rate of the viscous deformation and $\tilde{\mathbf{W}}_{v}$ is a skew-symmetric tensor representing the rate of stretching and spin, respectively. We make the intermediate state unique by prescribing $\tilde{\mathbf{W}}_{v} = \mathbf{0}$. The rate of the viscoelastic flow is described by
\begin{align}\label{eq: d_v}
	\tilde{\mathbf{D}}_{v} & =\ \frac{\dot{\varepsilon }_{v}}{\tau_{neq}}\ \text{dev}\left[ \boldsymbol{\sigma}_{neq}^{'}\right]
\end{align}
where $\tau_{neq} = \parallel \text{dev}[ \boldsymbol{\sigma}_{neq}] \parallel_{F}$ represents the Frobenius norm of the driving stress, $\dot{\varepsilon }^{v}$ is the viscous flow and $\boldsymbol{\sigma}_{neq}^{'} = \mathbf{R}_{e}^{T} \boldsymbol{\sigma}_{neq} \mathbf{R}_{e}$ represents the stress acting on the viscous component in its relaxed configuration. The viscous flow is defined by the Argon model
\begin{align}\label{eq: argon}
	\dot{\varepsilon}_{v} & = \dot{\varepsilon }_{0} \text{exp}\ \left[\frac{\Delta H}{k_{b} T} \ \left(\left(\frac{\tau_{neq}}{\tau_{0}}\right)^{m} -1\right)\right],
\end{align}
where $k_{b}$, $\dot{\varepsilon}_{0}$, $\Delta H$ and $\tau _{0}$ are the Boltzmann constant, a pre-exponential factor, the activation energy and the athermal yield stress. Recent models~\cite{poulain2014finite} have shown a better agreement with experimental data by using the exponential factor $m$ as a material parameter and therefore reconcile the moisture- or temperature dependency of the stiffness with the softening of the viscoelastic flow. We also modify the athermal yield stress of the argon model and propose a nonlinear behavior of the athermal yield stress driven by the local chain stretch $\Lambda_{chain}$, in contrast to the linear modification proposed by~\cite{qi2005stress}. 

It can be shown that the time derivative of $\dot{\bar{\mathbf{F}}}_{v}$ can be derived from \eqref{eq: d_v} and \eqref{eq: l_v} as follows:
\begin{align}\label{eq: fv_dot}
	\dot{\bar{\mathbf{F}}}_{v} & =\bar{\mathbf{F}}_{e}^{-1} . \dot{\varepsilon}_{v}\frac{\text{dev}\left[ \boldsymbol{\sigma}_{neq}^{'}\right]}{\tau_{neq}} . \bar{\mathbf{F}}_{ve}.
\end{align}

Similarly, the total velocity gradient of the overall network, $\bar{\mathbf{L}} = \dot{\bar{\mathbf{F}}}(\bar{\mathbf{F}})^{-1}$, can be expanded to the following:
\begin{align}
	\bar{\mathbf{L}} & = \bar{\mathbf{L}}_{ve} + \bar{\mathbf{L}}_{ve} . \bar{\mathbf{L}}_{vp} . \bar{\mathbf{F}}_{ve}^{-1} = \bar{\mathbf{L}}_{ve} + \tilde{\mathbf{L}}_{vp}.
\end{align}

Again, we consider the viscoplastic velocity gradient to be additively decomposed into the symmetric rate of stretching and the skew-symmetric rate of spinning:
\begin{align}
	\tilde{\mathbf{L}}_{vp} & = \dot{\bar{\mathbf{F}}}_{vp} . \bar{\mathbf{F}}_{vp}^{-1} = \tilde{\mathbf{D}}_{vp} + \tilde{\mathbf{W}}_{vp},
\end{align}
and we take $\tilde{\mathbf{W}}_{vp} = \mathbf{0}$ again leading to:
\begin{align}\label{eq: d_vp}
	\tilde{\mathbf{D}}_{vp} & = \frac{\dot{\varepsilon }_{vp}}{\tau_{tot}} \ \text{dev}\left[ \boldsymbol{\sigma}\right],
\end{align}
where $\text{dev}\left[ \boldsymbol{\sigma}\right]$ is the deviatoric part of the total stress and $\tau_{tot} = \parallel \text{dev}\left[ \boldsymbol{\sigma}\right]\parallel_{F}$.% is the Frobenius norm of the total deviatoric stress.
To characterize the viscoplastic flow $\dot{\varepsilon}_{vp}$, we implement a simple phenomenological representation as follows:
\begin{align}\label{eq: viscoplastic}
	\dot{\varepsilon }_{vp} & =\begin{cases}
		0 & \tau _{tot} < \ \sigma_{0}\\
		a( \epsilon \ -\ \epsilon_{0})^{b}\dot{\epsilon } & \tau_{tot} \ \geq \ \sigma_{0}
	\end{cases}, \ 
\end{align}
where $a, b$ and $\sigma_{0}$ are material parameters. $\epsilon_{0}$ is the stress at which the viscoplastic flow is activated, represented by the Frobenius norm of the Green strain tensor $\parallel \mathbf{E}\parallel_{F}$, which is derived from the deformation gradient:
\begin{align}\label{eq: e}
	\mathbf{E} & = \frac{1}{2} \ ( \mathbf{F}^{T} . \mathbf{F} - \mathbf{I}),
\end{align}
and $\dot{\epsilon}$ is the strain rate of the effective strain $\parallel \mathbf{E} \parallel_{F}$, thus introducing a simple strain-rate dependency of the viscoplastic flow. Analogous to~\ref{eq: fv_dot}, the time derivative of the viscoplastic deformation gradient is given by
\begin{align}\label{eq:fvp_dot}
	\dot{\bar{\mathbf{F}}}_{vp} & = \bar{\mathbf{F}}_{ve}^{-1} . \dot{\varepsilon}_{vp} \frac{\text{dev}\left[ \boldsymbol{\sigma}_{tot}\right]}{\tau_{tot}} . \bar{\mathbf{F}},
\end{align}
characterizing the rate kinematics of the viscoplastic flow. We obtain the viscous and viscoplastic deformation gradients at the end of a time increment using the Euler backward time integration. %The step-by-step procedure is presented in~\ref{sec:fem}.

In Eqs.~\ref{eq:fvp_dot} and \ref{eq: fv_dot}, the midpoint method is used to numerically obtain the inelastic deformation gradient at the end of a time increment, i.e.,
\begin{eqnarray}
	\bar{\boldsymbol{F}}_{v/vp}^{t+\frac{dt}{2}} & = & \bar{\boldsymbol{F}}_{v/vp}^{t}+\frac{dt}{2}\dot{\bar{\boldsymbol{F}}}_{v/vp}^{t},\label{eq:midpoint1}
\end{eqnarray}
\begin{eqnarray}
	\bar{\boldsymbol{F}}_{v/vp}^{t+dt} & = & \bar{\boldsymbol{F}}_{v/vp}^{t}+dt\dot{\bar{\boldsymbol{F}}}_{v/vp}^{t+\frac{dt}{2}}.\label{eq:midpoint2}
\end{eqnarray}

To calculate the elastic deformation gradient at the midpoint, it is also required to find the total deformation gradient at the midpoint. This is done by taking the average of the deformation gradient at the start and end of the increment
\begin{eqnarray}
	\bar{\boldsymbol{F}}^{t+\frac{dt}{2}} & = & \frac{\bar{\boldsymbol{F}}^{t}+\bar{\boldsymbol{F}}^{t+dt}}{2}.\label{eq:midpoint3}
\end{eqnarray}

\subsection{Phenomenological Viscoelastic-viscoplastic model coupled with a phase-field description}
\label{subsec:viscomodel}

Following the additive decomposition of the free energy proposed in~\cite{ambati2016phase}, the overall free energy of the material can be decomposed into an equilibrium $\psi_{eq}$ , a non-equilibrium $\psi_{neq}$ and a volumetric part $\psi_{vol}$ as
\begin{eqnarray}
	\psi\left(\bar{\boldsymbol{B}_{ve}},\bar{\boldsymbol{B}}_{e},J,\phi\right) & = & g\left(\phi\right)\psi_{0}^{+}\left(\bar{\boldsymbol{B}_{ve}},\bar{\boldsymbol{B}}_{e},J\right)+\psi_{0}^{-}\left(\bar{\boldsymbol{B}_{ve}},\bar{\boldsymbol{B}}_{e},J\right),\label{eq:free_energy_decomp}
\end{eqnarray}
where
\begin{eqnarray}
	\psi_{0}^{+} & = & \psi_{eq}\left(\bar{\boldsymbol{B}_{ve}}\right)+\psi_{neq}\left(\bar{\boldsymbol{B}}_{e}\right)+H\left(J-1\right)\psi_{vol}\left(J\right),\label{eq:psi_p}
\end{eqnarray}
and
\begin{eqnarray}
	\psi_{0}^{-} & = & \left(1-H\left(J-1\right)\right)\psi_{vol}\left(J\right).\label{eq:psi_m}
\end{eqnarray}

The Heaviside step function is defined as
\begin{eqnarray}
	H\left(x\right) & = & \begin{cases}
		0, & x<0\\
		1, & x\geq0
	\end{cases}.\label{eq:Hfun}
\end{eqnarray}

The energetic degradation function $g\left(\phi\right)$ captures the evolution of the strain energy versus the phase-field parameter and satisfies the following conditions
\begin{eqnarray}
	g\left(0\right) & = & 1,\:g\left(1\right)=0,\:g^{\prime}\left(\phi\right)\leq0\:\textrm{and}\:g^{\prime}\left(1\right)=0.\label{eq:g_cond}
\end{eqnarray}

The conditions prescribe a monotonic decreasing behavior during the fracture evolution. To prevent crack propagation under compression, the volumetric strain energy does not change when $J<1$ in Eqs.~\eqref{eq:psi_p} and \eqref{eq:psi_m}.

Here, the equilibrium $\psi_{eq}$ and non-equilibrium $\psi_{neq}$ parts of the free energy are defined by the neo-Hookean hyperelastic model as
\begin{eqnarray}
	\rho_{0}\psi_{eq} & = & \frac{1}{2}\mu_{eq}\left(\nu_{np},\Theta,w_w\right)\left(I_{1}\left(\bar{\boldsymbol{B}_{ve}}\right)-3\right),\label{eq:psi_eq}
\end{eqnarray}
and
\begin{eqnarray}
	\rho_{0}\psi_{neq} & = & \frac{1}{2}\mu_{neq}\left(\nu_{np},\Theta,w_w\right)\left(I_{1}\left(\bar{\boldsymbol{B}}_{e}\right)-3\right),\label{eq:psi_neq}
\end{eqnarray}
where $I_{1}\left(\cdot\right)=\textrm{tr}\left[\cdot\right]$ is the first invariant of the tensor. The material parameters $\mu_{eq}$ and $\mu_{neq}$ depend on temperature, BNP volume fraction $\nu_{np}$ and water content $w_w$
\begin{eqnarray}
	\mu_{eq}\left(\nu_{np},\Theta,w_w\right) & = & X\left(\nu_{np},w_w\right)\mu_{eq}^{0}\left(\Theta\right),\label{eq:mu}
\end{eqnarray}
\begin{eqnarray}
	\mu_{neq}\left(\nu_{np},\Theta,w_w\right) & = & X\left(\nu_{np},w_w\right)\mu_{neq}^{0}\left(\Theta\right).\label{eq:mu_e}
\end{eqnarray}

Assuming that BNPs are well-dispersed rigid particles in the epoxy matrix, the Guth–Gold model is adopted by which the effective stiffness of particle-filled solids is obtained by $\left\langle E\right\rangle =XE_{m}$~\cite{guth1945theory}. The amplification factor $X$ is typically a function of fillers' volume fraction and distribution. So far, some attempts of various levels of sophistication have been conducted to incorporate the effect of particle/matrix interactions on the effective modulus of polymer composites. Most of these models suggest a polynomial series expansion for the amplification factor. Here, a modified Guth–Gold model is adopted to account for uniformly distributed nanoparticles and moisture content as follows.
\begin{eqnarray}
	X & = & \left(1-9.5w_w+0.057 w_{w}^{2}\right) \left(1+3.5\nu_{np}+18\nu_{np}^{2}\right).\label{eq:X}
\end{eqnarray}

In Eq.~\eqref{eq:X}, the effect of moisture content on the material behavior is taken into account based on experimental data. The modified amplification factor proposed in this work is a first step to capture the stress-strain behavior of BNP/epoxy nanocomposites under hygrothermal conditions. Furthermore, to consider the effect of temperature on the material properties, a modified Kitagawa model proposed by Unger et al.~\cite{unger2020effect} is utilized with the following equations:
\begin{eqnarray}
	\mu_{eq}^{0} & = & \mu_{eq,ref}^{0}\left(2-\exp\left[\alpha\left(\varTheta-\varTheta_{ref}\right)\right]\right),\label{eq:mu_0}
\end{eqnarray}
\begin{eqnarray}
	\mu_{neq}^{0} & = & \mu_{neq,ref}^{0}\left(2-\exp\left[\alpha\left(\varTheta-\varTheta_{ref}\right)\right]\right).\label{eq:mu_e0}
\end{eqnarray}

The volumetric part of the free energy $\psi_{vol}$ is also defined by
\begin{eqnarray}
	\rho_{0}\psi_{vol} & = & \frac{1}{2}k_{v}\left(\frac{J_{m}^{2}-1}{2}-\ln\left[J_{m}\right]\right)^{2},\label{eq:psi_vol}
\end{eqnarray}
where the bulk modulus is assumed to be $k_{v}=\left(2-\exp\left[\alpha\left(\varTheta-\varTheta_{ref}\right)\right]\right)X k^{0}_{v}$. The total stresses are then obtained from 
\begin{eqnarray}
		\begin{cases}
			\boldsymbol{\sigma}=g\boldsymbol{\sigma}_{dev}+\boldsymbol{\sigma}_{vol} & J<1,\\
			\boldsymbol{\sigma}=g\boldsymbol{\sigma}_{dev}+g\boldsymbol{\sigma}_{vol} & J\geq1 .
		\end{cases}\label{eq:stress_ve}
\end{eqnarray}
where
\begin{eqnarray}
		\boldsymbol{\sigma}_{dev}=J^{-1}\left(\mu_{eq}\bar{\boldsymbol{B}_{ve}}^{D}+\mu_{neq}\bar{\boldsymbol{B}}_{e}^{D}\right),
		\label{eq:stress_ve1}
\end{eqnarray}
and
\begin{eqnarray}
		\boldsymbol{\sigma}_{vol}=\frac{1}{2}k_{v}J_{\Theta}^{-1}\left(J_{m}-\frac{1}{J_{m}}\right)\boldsymbol{1}.
		\label{eq:stress_ve2}
\end{eqnarray}

The constitutive model is able to capture the nonlinear elasticitiy at finite deformation, the nonlinear viscoelastic behavior, viscoplastic flow because of stress driven chain sliding, and the effect of temperature and moisture content on the stress-strain relationship.

\section{Phase-field model at finite deformation}
\label{sec:phase_model}

To evaluate the predictive capability of the proposed constitutive model, we use the model to present a formulation of the PFM for nanoparticle/polymer composites. This section presents a variational phase-field formulation for quasi-brittle
fracture at finite deformation, the continuum mechanics incremental scheme and FE equations are derived to show the procedure of analysis.

\subsection{Problem field description}
\label{sec:problem_field}

The strong form of the boundary value problem in the referential configuration for the coupled displacement $\boldsymbol{u}$ and phase-field variable $\phi$ can be written as
\begin{eqnarray}
\mathbf{\nabla_{X}}.\boldsymbol{P}+\boldsymbol{B}& = &\mathbf{0}\textrm{   in  }\varOmega_{0} \label{eq:bvp_rconf1}\\
\boldsymbol{P}.\mathbf{N}& = &\mathbf{\bar{T}}\textrm{  on  } \Gamma_{0} \label{eq:bvp_rconf2}\\
\frac{G_{c}}{l_{0}}\phi-G_{c}l_{0}\mathbf{\nabla_{X}}.\left(\mathbf{\nabla_{X}}\phi.\mathbf{C}^{-1}\right)& = &-g^{\prime}\left(\phi\right)\mathcal{H}\textrm{   in  }\varOmega_{0}\label{eq:bvp_rconf3}\\
\mathbf{\nabla_{X}}\phi.\boldsymbol{N}& = &0\textrm{  in  }\Gamma_{0} \label{eq:bvp_rconf4},
\end{eqnarray}
from which the Euler–Lagrange equations in the spatial form are obtained as
\begin{eqnarray}
\mathbf{\nabla_{x}}.\boldsymbol{\sigma}+\boldsymbol{b}& = &\mathbf{0}\textrm{   in  }\varOmega_{t} \label{eq:bvp1}\\
\boldsymbol{\sigma}.\mathbf{n}& = &\mathbf{\bar{t}}\textrm{  on  }\Gamma_{t} \label{eq:bvp2}\\
\frac{G_{c}}{l_{0}}\phi-G_{c}l_{0}\mathbf{\nabla_{x}}.\mathbf{\nabla_{x}}\phi& = &-g^{\prime}\left(\phi\right)\mathcal{H}\textrm{  in  }\varOmega_{t} \label{eq:bvp3}\\
\mathbf{\nabla_{x}}\phi.\boldsymbol{n}& = &0\textrm{  in  }\Gamma_{t} \label{eq:bvp4},
\end{eqnarray}

where $\boldsymbol{P}$ and $\boldsymbol{\sigma}$ are the first Piola-Kirchhoff stress and the right Cauchy–Green deformation tensor, respectively, and $\boldsymbol{C}=\boldsymbol{F}^T\boldsymbol{F}$ is the right Cauchy–Green deformation tensor. The coupled equations in the the spatial configuration is formulated by $\boldsymbol{\sigma}=\boldsymbol{P}.\text{cof}\left(\mathbf{F}^{-1}\right)$, $\mathbf{\nabla_{X}}\left(.\right)=\mathbf{\nabla_{x}\left(.\right).\mathbf{F}}$ and $\boldsymbol{b}=J^{-1}\boldsymbol{B}$, where $\text{cof}\left(\mathbf{F}\right)=J\mathbf{F}^{-T}$.

In the equations above, $l_{0}$ is the length scale that controls the width of the diffuse crack, $\Gamma=\Gamma_{t}\cup\Gamma_{u}$, $\boldsymbol{B}$ and $\boldsymbol{b}$ respetively represent the vector of body forces in the referential and spatial configuration, $\boldsymbol{\mathit{N}}$ and $\boldsymbol{\mathit{n}}$ are respectively the outward unit normal vector on the boundary $\Gamma_0$ of the body $\Omega_0$ and the boundary $\Gamma_t$ of the body $\Omega_t$, and $\boldsymbol{T}$ and $\boldsymbol{t}$ are respectively the traction force in the referential and spatial configuration. 

To take into account the effect of BNP content on the fracture evolution in the nanocomposites, the energy release rate is taken to be $G_{c}=X G^{0}_{c}$. Following a similar approach proposed in~\cite{dean2020phase}, the crack driving force $\mathcal{H}$ is defined by
\begin{eqnarray}
	\mathcal{H}\left(t\right) & = & \langle  \underset{\tau\in\left[0,t\right]}{\max}\psi_{0}^{+} - \psi_{0}^{+^{c}}|_{\sigma=\sigma_{d}} \rangle_{+}.\label{eq:H}
\end{eqnarray}

Here, $\mathcal{H}$ ensures the positive evolution of the phase field variable, which prevents the healing of cracks. $\underset{\tau\in\left[0,t\right]}{\max}\psi_{0}^{+}$ is the maximum ever reached free energy, and $\psi_{0}^{+^{c}}|_{\sigma=\sigma_{d}}$ is an effective energy for damage initiation that triggers the activation of damage. Here, $\sigma=\parallel  \boldsymbol{\sigma}\parallel_{F}$, and $\sigma_{d}$ is a material parameter.

Also, a monotonically decreasing degradation function $g(\phi)$, satisfying conditions presented in Eq.~\ref{eq:g_fun}, is given by
\begin{eqnarray}
	g\left(\phi\right) & = & \left(1-\phi\right)^{2}+k,\label{eq:g_fun}
\end{eqnarray}
where $k$ is a small positive parameter introduced for ensuring the stability of the solution~\cite{miehe2010thermodynamically}.

%\subsection{Weak from formulation}
%\label{sec:fem}

To obtain the weak form of the governing equations, the weighted residual approach is used. Accordingly, Eqs.~\eqref{eq:bvp_rconf1} and \eqref{eq:bvp_rconf3} multiplied by weight functions and integrated over $\varOmega_0$. Using the divergence theorem and imposing the boundary conditions, the weak form of the governing equations is obtained as follows
\begin{eqnarray}
\int_{\varOmega_{0}}\boldsymbol{P}.\mathbf{\nabla_{X}}.\boldsymbol{\eta}_{u}\text{d}V-\int_{\varOmega_{0}}\rho_{0}\boldsymbol{B}.\boldsymbol{\eta}_{u}\text{d}V-\int_{\varOmega_{0}}\bar{\boldsymbol{T}}.\boldsymbol{\eta}_{u}\text{d}A & = & 0\quad\forall\quad\boldsymbol{\eta}_{u}\quad\in H_{0}^{1}\left(\varOmega\right), \label{eq:weak_rconf1}
\end{eqnarray}
and
\begin{eqnarray}
\int_{\varOmega_{0}}\left(g^{\prime}\left(\phi\right)\mathcal{H}\eta_{\phi}+\frac{G_{c}}{l_{0}}\phi\eta_{\phi}-G_{c}l_{0}\mathbf{\nabla_{X}}\phi.\mathbf{C}^{-1}.\mathbf{\nabla_{X}}\eta_{\phi}\right)\text{d}V & = & 0\quad\forall\quad\eta_{\phi}\in H_{0}^{1}\left(\varOmega\right). \label{eq:weak_rconf2}
\end{eqnarray}

The weak form in the spatial description can then be derived as
\begin{eqnarray}
	\int_{\varOmega_{t}}\boldsymbol{\sigma}.\mathbf{\nabla_{x}}.\boldsymbol{\eta}_{u}\text{d}v-\int_{\varOmega_{t}}\rho_{t}\boldsymbol{b}.\boldsymbol{\eta}_{u}\text{d}v-\int_{\varOmega_{t}}\bar{\boldsymbol{t}}.\boldsymbol{\eta}_{u}\text{d}a & = & 0\quad\forall\quad\boldsymbol{\eta}_{u}\in H_{0}^{1}\left(\varOmega\right), \label{eq:weak_u}
\end{eqnarray}
and
\begin{eqnarray}
	\int_{\varOmega_{t}}\left(J^{-1}g^{\prime}\left(\phi\right)\mathcal{H}\eta_{\phi}+J^{-1}\frac{G_{c}}{l_{0}}\phi\eta_{\phi}-J^{-1}G_{c}l_{0}\mathbf{\nabla_{x}}\phi.\mathbf{\nabla_{x}}\eta_{\phi}\right)\text{d}v & = & 0\quad\forall\quad\eta_{\phi}\in H_{0}^{1}\left(\varOmega\right). \label{eq:weak_phi}
\end{eqnarray}

The Eqs.~\eqref{eq:weak_u}~and~\eqref{eq:weak_phi} in terms of spatial arguments are formulated using $dv=JdV$ and $\boldsymbol{n}da=cof\left(\mathbf{F}\right)$.

\subsection{Consistent incremental-iterative scheme}
\label{sec:increment_scheme}

Assuming that only deformation-independent loads act on the body, Eqs.~\ref{eq:weak_rconf1} and \ref{eq:weak_rconf2} can then be expressed in terms of external and internal nodal forces as
\begin{eqnarray}
	\mathbf{r}^{u} & = & \mathbf{f}_{int}^{u}-\mathbf{f}_{ext}^{u}=\mathbf{0},\label{eq:res_rconf1}
\end{eqnarray}
\begin{eqnarray}
	r^{\phi} & = & f_{int}^{\phi}-f_{ext}^{\phi}=0, \label{eq:res_rconf2}
\end{eqnarray}
where
\begin{eqnarray}
	\mathbf{f}_{int}^{u} & = & \int_{\varOmega_{0}}\boldsymbol{P}.\mathbf{\nabla_{X}}.\boldsymbol{\eta}_{u}\text{d}V, \label{eq:fint_rconf_u}
\end{eqnarray}
\begin{eqnarray}
	\mathbf{f}_{ext}^{u} & = & \int_{\varOmega_{0}}\rho_{0}\boldsymbol{B}.\boldsymbol{\eta}_{u}\text{d}V-\int_{\varOmega_{0}}\bar{\boldsymbol{T}}.\boldsymbol{\eta}_{u}\text{d}A, \label{eq:fext_rconf_u}
\end{eqnarray}
\begin{eqnarray}
	\mathbf{f}_{int}^{\phi} & = & \int_{\varOmega_{t}}\left(J^{-1}g^{\prime}\left(\phi\right)\mathcal{H}\eta_{\phi}+J^{-1}\frac{G_{c}}{l_{0}}\phi\eta_{\phi}-J^{-1}G_{c}l_{0}\mathbf{\nabla_{x}}\phi.\mathbf{\nabla_{x}}\eta_{\phi}\right)\text{d}v, \label{eq:fint_rconf_phi}
\end{eqnarray}
\begin{eqnarray}
	\mathbf{f}_{ext}^{\phi} & = & \mathbf{0}, \label{eq:fext_rconf_phi}
\end{eqnarray}

By linearizing Eqs.~\eqref{eq:res_rconf1} and \eqref{eq:res_rconf2} at iteration $i + 1$ with respect to the previous iteration $i$, a consistent tangent stiffness is obtained as follows
\begin{eqnarray}
	\mathbf{r}_{i+1}^{u} & = & \mathbf{r}_{i}^{u}+\triangle\mathbf{r}^{u}=\mathbf{0},
\end{eqnarray}
\begin{eqnarray}
	r_{i+1}^{\phi} & = & r_{i}^{\phi}+\triangle r^{\phi},
\end{eqnarray}
where
\begin{eqnarray}
	\triangle\mathbf{r}^{u} & = & D_{u}\mathbf{r}_{i}^{u}.\triangle\mathbf{u}+D_{\phi}\mathbf{r}_{i}^{u}.\triangle\phi
\end{eqnarray}
and
\begin{eqnarray}
	\triangle r^{\phi} & = & D_{u}r_{i}^{\phi}.\triangle\mathbf{u}+D_{\phi}r_{i}^{\phi}.\triangle\phi.
\end{eqnarray}

The linearization of Eqs.~\eqref{eq:weak_u} and \eqref{eq:weak_phi} in spatial formulation can finally be obtained using a push fowrard of the linearized equation in the referential configuration
\begin{eqnarray}
	\int_{\varOmega_{t}}\left(\mathbf{\nabla_{x}}\triangle\mathbf{u}.\boldsymbol{\sigma}.\mathbf{\nabla_{x}}\boldsymbol{\eta}_{u}+\mathbf{\nabla_{x}^{s}}\boldsymbol{\eta}_{u}.\mathbf{c}.\mathbf{\nabla_{x}^{s}}\triangle\mathbf{u}\right)\text{d}v+\int_{\varOmega_{t}}\left(\mathbf{\nabla_{x}^{s}}\boldsymbol{\eta}_{u}.D_{\phi}\boldsymbol{\sigma}.\triangle\phi\right)\text{d}v & = & \mathbf{f}_{ext}^{u}-\mathbf{f}_{int,i}^{u}
\end{eqnarray}
and
\begin{eqnarray}
	\int_{\varOmega_{t}}\left(J^{-1}g^{\prime}\left(\phi\right)2\frac{\partial\mathcal{H}}{\partial\mathbf{g}}.\mathbf{\nabla_{x}}\triangle\mathbf{u}\eta_{\phi}\right)\text{d}v \notag\\
	+\int_{\varOmega_{t}}\left(J^{-1}g^{\prime\prime}\left(\phi\right)\mathcal{H}\triangle\phi\eta_{\phi}+J^{-1}\frac{G_{c}}{l_{0}}\triangle\phi\eta_{\phi}+J^{-1}G_{c}l_{0}\mathbf{\nabla_{x}}\triangle\phi.\nabla_{x}\eta_{\phi}\right)\text{d}v & = & f_{ext}^{\phi}-f_{int,i}^{\phi}
\end{eqnarray}

In the equations above, $\mathbf{\hat{c}} = J^{-1} \mathbf{c}=\frac{\partial\boldsymbol{S}}{\partial\boldsymbol{C}}$, where $\boldsymbol{S}$ is the second Piola-Kirchhoff stress. Also, $\frac{\partial(.)}{\partial\boldsymbol{g}}= \boldsymbol{F} \frac{\partial(.)}{\partial\boldsymbol{C}} \boldsymbol{F}^T$.

\subsection{Finite element formulation}

The linearized equilibrium and phase-field equations are finally summarized in the following system of equations:
\begin{eqnarray}
	\begin{bmatrix}\mathbf{K}_{i}^{uu} & \mathbf{K}_{i}^{u\phi}\\
		\mathbf{K}_{i}^{\phi u} & \mathbf{K}_{i}^{\phi\phi}
	\end{bmatrix}\begin{bmatrix}\delta\mathbf{u}_{i+1}\\
		\delta\mathbf{\boldsymbol{\phi}}_{i+1}
	\end{bmatrix} & = & \begin{bmatrix}\mathbf{\mathit{\mathbf{f}}}_{ext}^{u}\\
		\mathbf{\mathit{\mathbf{f}}}_{ext}^{\phi}
	\end{bmatrix}-\begin{bmatrix}\mathbf{\mathit{\mathbf{f}}}_{int,i}^{u}\\
		\mathbf{\mathit{\mathbf{f}}}_{int,i}^{\phi}
	\end{bmatrix}\label{eq:matrix_form}
\end{eqnarray}
where
%
%\begin{eqnarray}
%	\mathbf{K}_{i}^{uu} & = & \int_{\varOmega}\mathbf{B}_{u}^{T}\left(\frac{\partial\boldsymbol{\sigma}}{\partial\boldsymbol{\varepsilon}}\right)\mathbf{B}_{u}\text{d}\varOmega + \int_{\varOmega}\mathbf{B}_{u}^{T} \boldsymbol{\sigma} \mathbf{B}_{u}\text{d}\varOmega,\label{eq:kuu}
%\end{eqnarray}
\begin{eqnarray}
	\mathbf{K}_{i}^{uu} & = & \int_{\varOmega}\mathbf{B}_{u}^{T} \mathbf{\hat{c}} \mathbf{B}_{u}\text{d}\varOmega + \int_{\varOmega}\mathbf{B}_{u}^{T} \boldsymbol{\sigma} \mathbf{B}_{u}\text{d}\varOmega,\label{eq:kuu}
\end{eqnarray}
\begin{eqnarray}
	\mathbf{K}_{i}^{u\phi} & = & \int_{\varOmega}\mathbf{B}_{u}^{T}\left(\frac{\partial\boldsymbol{\sigma}}{\partial\phi}\right)\mathbf{N}_{\phi}\text{d}\varOmega,\label{eq:kuphi}
\end{eqnarray}
\begin{eqnarray}
	\mathbf{K}_{i}^{\phi u} & = & \int_{\varOmega}\mathbf{N}_{\phi}^{T}\left(J^{-1} g^{\prime}\frac{2\partial\mathcal{H}}{\partial\boldsymbol{g}}\right)\mathbf{B}_{u}\text{d}\varOmega,\label{eq:kphiu}
\end{eqnarray}
\begin{eqnarray}
	\mathbf{K}_{i}^{\phi\phi} & = & \int_{\varOmega}J^{-1}\left(\mathbf{N}_{\phi}^{T}\left(g^{\prime\prime}\mathcal{H}+\frac{G_{c}}{l_{0}}\right)\mathbf{N}_{\phi}+G_{c}l_{0}\mathbf{B}_{\phi}^{T}\mathbf{B}_{\phi}\right)\text{d}\varOmega,\label{eq:kphiphi}
\end{eqnarray}

In the equations above, the shape function matrices $\boldsymbol{N}_{u}$ and $\boldsymbol{N}_{\phi}$ interpolate the nodal values $\boldsymbol{u}$ and $\boldsymbol{\phi}$, respectively, and $\boldsymbol{B}_{u}$ and $\boldsymbol{B}_{\phi}$ are the gradient operators for the displacement and the nonlocal equivalent strain, respectively. The same shape functions interpolate the nodal values of the weight functions $\mathbf{\eta}_{u}$ and $\eta_{\phi}$.

In what follows, the two fields of the phase-field fracture problem are solved using a staggered algorithm proposed by Miehe et al.~\cite{miehe2010phase}. The staggered solution, which decouples the equilibrium and phase-field equations, has shown to be computationally efficient and robust through its broad application in the literature~\cite{molnar20172d,dean2020phase}. Accordingly, the coupling terms $\mathbf{K}_{i}^{u \phi }$ and $\mathbf{K}_{i}^{\phi u}$ are not taken into account.

\subsection{Consistent tangent moduli based on the Jaumann–Zaremba stress rate}
\label{sec:tangent_moduli}

In order to integrate the viscoelastic model into the incremental-iterative FE framework, the tangent modulus tensor $\mathbb{C}^{\sigma J}=\frac{\partial\boldsymbol{\sigma}}{\partial\boldsymbol{\varepsilon}}$ needs to be explicitly specified. However, a closed-form calculation of the tangent tensor is not a straightforward task. Here, we use an efficient numerical approximation of the tangent moduli proposed by Sun et al.~\cite{sun2008numerical}. In this approach, by perturbing the deformation gradient, the tangent moduli for the Jaumann rate of the Cauchy stress are accurately approximated by a forward difference of the Cauchy stresses. The Jaumann rate of the Cauchy stress can be expressed as
\begin{eqnarray}
	\overset{\bigtriangledown}{\boldsymbol{\sigma}} & = &  
	\dot{\boldsymbol{\sigma}}-\boldsymbol{W}\boldsymbol{\sigma}-\boldsymbol{\sigma}\boldsymbol{W}^{T}=\mathbb{C}^{\sigma J}:\boldsymbol{D}\label{eq:jaumann_stress_rate}
\end{eqnarray}

The linearized incremental form of Eq.~\eqref{eq:jaumann_stress_rate} is then obtained from
\begin{eqnarray}
	\Delta\boldsymbol{\sigma}-\Delta\boldsymbol{W}\boldsymbol{\sigma}-\boldsymbol{\sigma}\Delta\boldsymbol{W}^{T} & = & \mathbb{C}^{\sigma J}:\Delta\boldsymbol{D}\label{eq:del_sigma}
\end{eqnarray}

To numerically calculate components of $\mathbb{C}^{\sigma J}$, Eq.~\eqref{eq:del_sigma} is perturbed by applying small perturbations to components of $\Delta\boldsymbol{D}$ and $\Delta\boldsymbol{W}$ tensors. Here, $\Delta\boldsymbol{W}_{ij}$ and $\Delta\boldsymbol{D}_{ij}$ tensors with perturbed ($i$,$j$) components are expressed as
\begin{eqnarray}
	\Delta\boldsymbol{W}_{ij} & = & \frac{1}{2}\left(\Delta\boldsymbol{F}_{ij}\boldsymbol{F}^{-1}-\left(\Delta\boldsymbol{F}_{ij}\boldsymbol{F}^{-1}\right)^{T}\right),\label{eq:del_W}
\end{eqnarray}
and
\begin{eqnarray}
	\Delta\boldsymbol{D}_{ij} & = & \frac{1}{2}\left(\Delta\boldsymbol{F}_{ij}\boldsymbol{F}^{-1}+\left(\Delta\boldsymbol{F}_{ij}\boldsymbol{F}^{-1}\right)^{T}\right),\label{eq:del_D}
\end{eqnarray}
where the corresponding perturbed $\Delta\boldsymbol{F}_{ij}$ is obtained from perturbing its ($i$,$j$) component as~\cite{miehe1996numerical}
\begin{eqnarray}
	\Delta\boldsymbol{F}_{ij} & = & \frac{\epsilon}{2}\left(e_{i}\otimes e_{j}\boldsymbol{F}+e_{j}\otimes e_{i}\boldsymbol{F}\right).\label{eq:del_F}
\end{eqnarray}

By substituting Eq.~\eqref{eq:del_F} into Eqs.~\eqref{eq:del_W}~and~\eqref{eq:del_D}, we have
\begin{eqnarray}
	\Delta\boldsymbol{W}_{ij} & = & \boldsymbol{0},\label{eq:del_W2}
\end{eqnarray}
\begin{eqnarray}
	\Delta\boldsymbol{D}_{ij} & = & \frac{\epsilon}{2}\left(e_{i}\otimes e_{j}+e_{j}\otimes e_{i}\right).\label{eq:del_D2}
\end{eqnarray}

It is noteworthy that $\Delta\boldsymbol{D}$ has six independent components due to its symmetry. Therefore, the choice of ($i$,$j$) would be (1,1), (2,2), (3,3), (1,2), (1,3), and (2,3). The perturbed deformation gradient $\hat{\boldsymbol{F}}_{ij}$ can then be written as
\begin{eqnarray}
	\hat{\boldsymbol{F}}_{ij} & = & \boldsymbol{F}+\Delta\boldsymbol{F}_{ij}\label{eq:F_hat}
\end{eqnarray}

Using Eq.~\eqref{eq:F_hat}, $\Delta\boldsymbol{\sigma}$ is approximated by the forward difference of the perturbed and unperturbed Cauchy stresses
\begin{eqnarray}
	\Delta\boldsymbol{\sigma} & \approx & \boldsymbol{\sigma}\left(\hat{\boldsymbol{F}}_{ij}\right)-\boldsymbol{\sigma}\left(\boldsymbol{F}\right).\label{eq:del_sigma2}
\end{eqnarray}

Substituting Eqs.~\eqref{eq:del_W2}, \eqref{eq:del_D2} and \eqref{eq:del_sigma2} into Eq.~\eqref{eq:del_sigma} gives
\begin{eqnarray}
	\boldsymbol{\sigma}\left(\hat{\boldsymbol{F}}_{ij}\right)-\boldsymbol{\sigma}\left(\boldsymbol{F}\right) & \approx & \mathbb{c}_{ij}^{\sigma J}:\frac{\epsilon}{2}\left(e_{i}\otimes e_{j}+e_{j}\otimes e_{i}\right).\label{eq:sigma_hat}
\end{eqnarray}

Using Eq.~\eqref{eq:sigma_hat}, the numerical approximation of the tangent moduli is finally obtained as
\begin{eqnarray}
	\mathbb{C}_{ij}^{\sigma J} & = & \frac{1}{\epsilon}\left[\boldsymbol{\sigma}\left(\hat{\boldsymbol{F}}_{ij}\right)-\boldsymbol{\sigma}\left(\boldsymbol{F}\right)\right]\label{eq:CJ}
\end{eqnarray}
where $\mathbb{C}_{ij}^{\sigma J}$ represents the components of the tangent modulus tensor $\mathbb{C}^{\sigma J}$ calculated by the perturbation of $\Delta\boldsymbol{F}_{ij}$. Table~\ref{table:tangent_algorithem} summarizes a step-by-step algorithm adopted from \cite{ostwald2019implementation} to compute the mechanical tangent and solve the system of non-linear equations.

\begin{table}[h]
		\centering
		\caption{Summary of step-by-step algorithm used to compute the tangent moduli and solve the system of non-linear equations.}
		\label{table:tangent_algorithem}
		\begin{tabularx}{\textwidth}{|X|l}
			%\begin{tabularx}{|l|}
			\hline	
			1. Define a perturbation parameter: $\epsilon=10^{-8}$.\\
			2. Compute the right Cauchy-Green tensor: $\mathbf{C}=\mathbf{F}^T \mathbf{F}$. \\
			3. Calculate the Cauchy stress tensor (i.e., $\boldsymbol{\sigma}$) using equations presented in Sec.~\ref{sec:const}. \\
			4. \textbf{for} $k=1 ... 3$ \textbf{do} \\
			5. \quad \textbf{for} $l=1 ... 3$ \textbf{do} \\
			6. \quad \quad Initialize $\hat{\boldsymbol{F}} = \boldsymbol{F}$ \\ 
			7. \quad \quad Perturb $\hat{\boldsymbol{F}}_{kl} =\hat{\boldsymbol{F}}_{kl} + \epsilon/2$. \\
			8. \quad \quad Perturb $\hat{\boldsymbol{F}}_{lk} =\hat{\boldsymbol{F}}_{lk} + \epsilon/2$. \\	
			9. \quad \quad Calculate the corresponding perturbed stress response $\Delta\boldsymbol{\sigma}$ using Eq.~\eqref{eq:del_sigma2}. \\
			10. \quad \quad \textbf{for} $i=1 ... 3$ \textbf{do} \\	
			11. \quad \quad \quad \textbf{for} $j=1 ... 3$ \textbf{do} \\	
			12. \quad \quad \quad \quad Compute and store $\mathbb{C}_{ijkl}^{\sigma J}$ and $\mathbb{C}_{ijlk}^{\sigma J}$ based on the Jaumann stress rate of the Cauchy stress as presented in Eq.~\eqref{eq:CJ}. \\
			13. \quad \quad \quad end \\
			14. \quad \quad end \\
			15. \quad end \\
			16. end \\
			17. Determine the material tangent: $\mathbb{C}^{\sigma J}$. \\
			18. Store the tangent tensor in Voigt-type matrix. \\
			19. Solve the system of equations presented in Eq.~\eqref{eq:matrix_form} using a staggered algorithm~\cite{miehe2010phase}. \\
			\hline		
	\end{tabularx}
\end{table}

\section{Fracture experiments and numerical simulations}
\label{sec:results}

This section aims to identify and validate the proposed PFM's parameters. Firstly, we compare the numerical results of dogbone tests of BNP/epoxy samples with experimental data. Then, we study the influence of temperature on the fracture evolution. Lastly, we qualitatively evaluate the model's ability to predict fracture patterns using the well-known single-edge notched tests.

\subsection{Experiments}
\label{subsec:experiments}

\begin{figure}[h]
	\centering
	\tikzset{every picture/.style={line width=0.75pt}} %set default line width to 0.75pt        
	\begin{tikzpicture}[x=0.75pt,y=0.75pt,yscale=-1,xscale=1]
		\node[inner sep=0pt]  at (0.0,0.0)	{\includegraphics[scale=1]{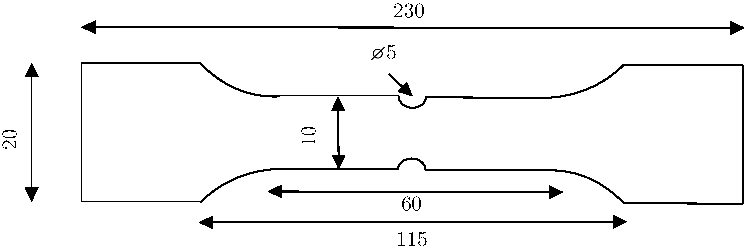}};
	\end{tikzpicture}
	\caption{Planar dimensions of the specimen for conditioning and mechanical loading-unloading tests with a thickness of 2.3 mm. All dimensions are in millimeters.}
	\label{fig: spec}
\end{figure}

The samples used for conditioning and mechanical tests are obtained from the panels shown in Figure~\ref{fig: spec}. In epoxy systems, necking in the tension direction is not a material property but rather a structural instability~\cite{poulain2014finite}. Therefore, a notch is introduced to mitigate the effects of material imperfections and necking on the yield. The samples are conditioned at 60 ◦C and 85\% relative humidity until they reach the saturated state at a moisture concentration of 1.0\% for the neat epoxy system and 1.2\% for the BNPs reinforced epoxy. The total conditioning time was 115 days. Finally, the samples are subjected to mechanical loading-unloading tests according to the DIN EN ISO 527–2 testing standard. An extensometer is used to measure the specimen's elongation, and a load rate of 1 mm/min is applied. Before the final loading cycle that led to failure, six cycles were performed by loading the specimen to a specific amplitude and unloading it until the loading force reached zero.

\subsection{Simulations}

\begin{figure}[h]
	\centering
		\begin{tikzpicture}[scale=2.5]
			%\clip (-1.9,-0.8) rectangle (1.8, 0.8);
			\node[inner sep=0pt]  at (0.0,0.0)	{\includegraphics[scale=1]{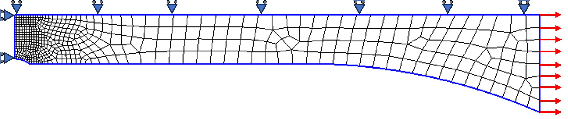}};
		\end{tikzpicture}
		\caption{Loading and boundary conditions imposed on quarter of the specimen because of symmetry, and a two-dimensional FE model composed of 1563 Q4 elements.}
		\label{fig:db_mesh}
\end{figure}

The specimen's symmetry allows for a more efficient FE analysis. By using the symmetries and analyzing a quarter of the specimen, we can obtain a complete solution of the entire model with less computational cost. As illustrated in Fig.~\ref{fig:db_mesh}, the model considers the double symmetry of the sample at the mid-length and width. The uppermost side's vertical displacement and the left side's horizontal displacement of the specimen are restrained, while a horizontal displacement is applied to the right side of the specimen, as shown in the figure.

Fig.~\ref{fig:db_mesh} shows the corresponding mesh with 1563 four-noded quadrilateral (Q4) elements. Due to high-stress concentrations in the right part of the model, the mesh size is refined to around 0.1 mm in the area, while the length scale parameter is set to 0.5 mm. The load is applied via an imposed displacement at a constant deformation rate of $\dot{u}=1$~mm/min and $\Theta=296$~K. The following simulations are performed under plane strain conditions, with the load applied through constant displacement increments of $\Delta u=1\times10^{-4}$~mm. The simulations have been performed using an in-house parallel MATLAB code.

Table~\ref{tab:param} provides the identified parameters of the PFM. The table shows that the material parameters linked with equilibrium, non-equilibrium, and viscoelastic responses, the modified Kitagawa model, moisture swelling, the energy release rate of the neat epoxy, and the length scale have been obtained from the authors' previous studies~\cite{arash2023effect,BAHTIRI2023116293}. Additionally, the remaining material parameters have been calibrated to achieve the best fit to the experimental data. The parameters are validated using experimental data for the epoxy system with 0\% and 5\% wt BNPs at dry and saturated conditions. The mass density of the neat epoxy and BNP are 1.2 and 3.0~g/cc, respectively. Accordingly, the corresponding volume fraction can be calculated to be $\nu_{np}=0.0215$.

\begin{table}[h]
	\centering
	\caption{Materials parameters of the PFM model.} \label{tab:param}
	\begin{footnotesize}
		\begin{tabular}{ccccc}
			\toprule 
			& Parameter & Value & Equation & References \\
			\midrule 
			\multicolumn{1}{l}{Equilibrium shear modulus} & $\mu_{eq}^{0}(\text{MPa})$ &760& \ref{eq:stress_ve1}& \cite{BAHTIRI2023116293}  \\
			\multicolumn{1}{l}{Non-equilibrium shear modulus}  & $\mu _{neq}^{0}(\text{MPa})$ & 790 &\ref{eq:stress_ve1} & \cite{BAHTIRI2023116293} \\
			\multicolumn{1}{l}{Volumetric bulk modulus} & $\displaystyle \kappa_{v} (\text{MPa})$ & 1154 &\ref{eq:stress_ve2}& \cite{BAHTIRI2023116293} \\
			\multicolumn{1}{l}{Viscoelastic dashpot} & $\displaystyle \dot{\varepsilon }_{0}\left(s^{-1}\right)$ & 1.0447 x 10\textsuperscript{12} &\ref{eq: argon} & \cite{BAHTIRI2023116293} \\
			& $\Delta H(J)$ & 1.977 x 10\textsuperscript{-19} &\ref{eq: argon}& \cite{BAHTIRI2023116293}  \\
			& $m$ & 0.657 &\ref{eq: argon}& \cite{BAHTIRI2023116293}\\
			& $\tau_{0}$ & 40 &\ref{eq: argon}&  \\
			\multicolumn{1}{l}{Viscoplastic dashpot} & $a$ & 0.1 & \ref{eq: viscoplastic} &  \\
			& \"b & $22 \omega_w + 0.8$ &\ref{eq: viscoplastic}&  \\
			& $\sigma_{0}(\text{MPa})$ & $30 X$ &\ref{eq: viscoplastic}&  \\
			\multicolumn{1}{l}{Energy release rate} & $G^{0}_{c}$ (N/mm) & 190 &\ref{eq:bvp3}& \cite{arash2023effect} \\
			\multicolumn{1}{l}{Length scale parameter} & $l_{0}$ (mm) & 0.5 &\ref{eq:bvp3}& \cite{arash2023effect} \\	
			\multicolumn{1}{l}{$\sigma_{d}$} & $\sigma_{d}$ (MPa) & $55X$  &\ref{eq:H}& \cite{arash2023effect} \\	
			\multicolumn{1}{l}{Moisture swelling coefficient} & $\displaystyle a_{w}$ & 0.039 &\ref{eq:Jw} & \cite{arash2023effect} \\
			\bottomrule
		\end{tabular}
	\end{footnotesize}
\end{table}

\begin{figure}[!h]
	\centering
	\begin{subfigure}[b]{0.5\textwidth}
			\centering
			\begin{tikzpicture}
			\node[inner sep=0pt]  at (0.0,0.0)	{\includegraphics[scale=1]{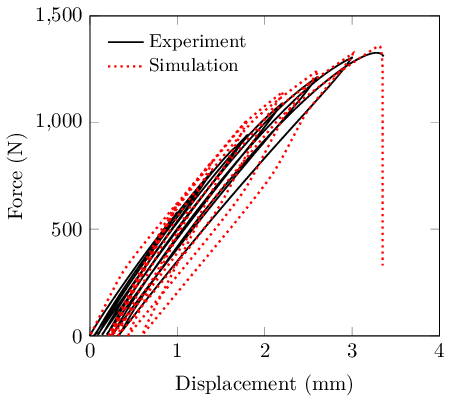}};
			\end{tikzpicture}
			\caption{}
			\label{fig:force_disp_np0_w0}
		\end{subfigure}%
	\begin{subfigure}[b]{0.5\textwidth}
		\centering
		\begin{tikzpicture}
			\node[inner sep=0pt]  at (0.0,0.0)	{\includegraphics[scale=1]{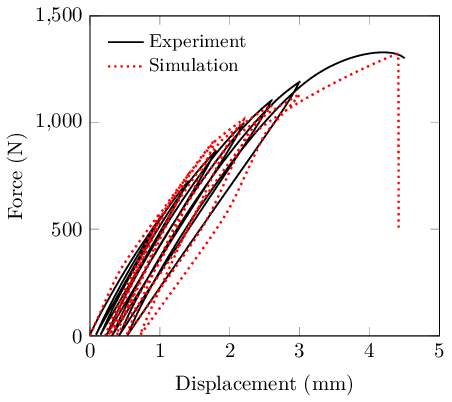}};			
		\end{tikzpicture}
		\caption{}
		\label{fig:force_disp_np0_w01}
	\end{subfigure}%
	\caption{Effect of moisture on the force–displacement response of the epoxy at room temperature and the deformation rate of 1~mm/min: (a) dry sample, and (b) saturated sample.}
	\label{fig:db_np0}
\end{figure}

The resulting cyclic loading-unloading force-displacement curves of the dogbone tests for dry and saturated epoxy samples are compared with experimental data in Fig.~\ref{fig:db_np0}. The experimental-numerical comparison demonstrates a good agreement between the experimental data and the numerical predictions along the whole evolution of displacement. It evidences the capability of the proposed PFM in predicting the evolution of damage and viscoplasticy mechanisms in the epoxy polymer under cyclic loading. Comparing the experimental data in Figs.~\ref{fig:force_disp_np0_w0} and \ref{fig:force_disp_np0_w01} show that the ductility of the epoxy increases at the presence of moisture content.

\begin{figure}[!h]
	\centering
	\begin{subfigure}[b]{0.33\linewidth}
		\centering
		\begin{tikzpicture}[scale=2]
			\node[inner sep=0pt]  at (0.0,0.0)	{\includegraphics[scale=1]{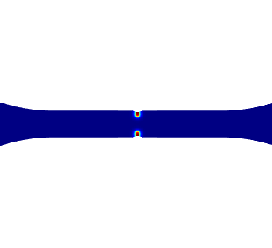}};
		\end{tikzpicture}
		\caption{ }
		\label{fig:fracture_db1}
	\end{subfigure}%
	\begin{subfigure}[b]{0.33\linewidth}
		\centering	
		\begin{tikzpicture}[scale=2]
			\node[inner sep=0pt]  at (0.0,0.0)	{\includegraphics[scale=1]{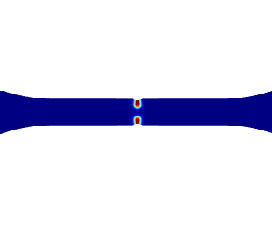}};
		\end{tikzpicture}
		\caption{}			
		\label{fig:fracture_db2}
	\end{subfigure}	
	\begin{subfigure}[b]{0.33\linewidth}
	\centering	
	\begin{tikzpicture}[scale=2]
			\node[inner sep=0pt]  at (0.0,0.0)	{\includegraphics[scale=1]{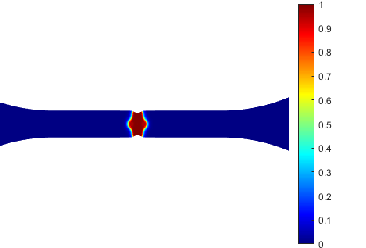}};		
	\end{tikzpicture}
	\caption{}			
	\label{fig:fracture_db3}			
	\end{subfigure}
	\caption{Contour plots of damage in the dry epoxy sample for imposed displacements of (a) 3.32~mm, (b) 3.34~mm, and (c) 3.35~mm.}
	\label{fig:db_contours}
\end{figure}

The evolution of the crack phase-field at imposed displacements of 3.32, 3.34 and 3.35~mm is also illustrated in Figs.~\ref{fig:fracture_db1}-\ref{fig:fracture_db3}. 

\begin{figure}[h]
	\centering
	\begin{subfigure}[b]{0.5\textwidth}
		\centering
		\begin{tikzpicture}
			\node[inner sep=0pt]  at (0.0,0.0)	{\includegraphics[scale=1]{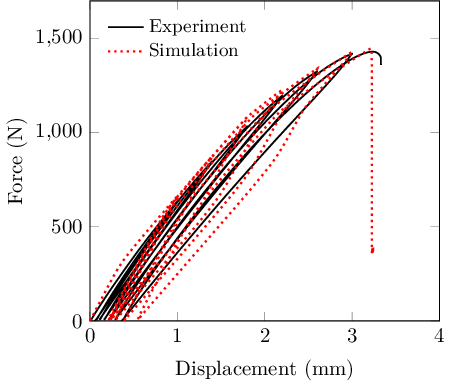}};
		\end{tikzpicture}
		\caption{}
		\label{fig:force_disp_np05_w0}
	\end{subfigure}%
	\begin{subfigure}[b]{0.5\textwidth}
		\centering
		\begin{tikzpicture}
			\node[inner sep=0pt]  at (0.0,0.0)	{\includegraphics[scale=1]{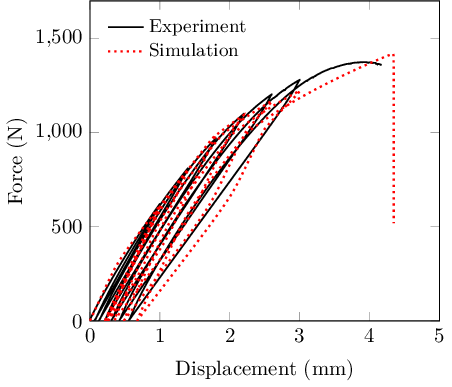}};			
		\end{tikzpicture}
		\caption{}
		\label{fig:force_disp_np05_w1}
	\end{subfigure}%
	\caption{Effect of moisture on the force–displacement response of BNP(5~\%wt)/epoxy at room temperature and the deformation rate of 1~mm/min: (a) dry sample, and (b) saturated sample.}
	\label{fig:db_np05}
\end{figure}

The results presented in Figs.~\ref{fig:force_disp_np05_w0} and \ref{fig:force_disp_np05_w1} demonstrate a satisfactory agreement between the numerical predictions and experimental data for the epoxy system with 5\% weight percent of BNPs in both dry and saturated conditions. Although the proposed PFM model can reasonably predict the highly nonlinear viscoelastic-viscoplastic behavior of the material, it has some limitations when compared to the experimental data. The deviation from the experimental results could be attributed to the formulation of the constitutive model itself as well as the unique set of material parameters used. Nevertheless, the amplification approach employed in this study to account for the effects of moisture and BNP contents has resulted in realistic numerical predictions. Additionally, by calibrating the viscoplastic flow and modifying the crack driving force, we were able to reasonably predict the evolution of plasticity and damage in the nanocomposites.

\begin{figure}[!h]
	\centering
	\begin{subfigure}[b]{0.5\textwidth}
		\centering
		\begin{tikzpicture}
			\node[inner sep=0pt]  at (0.0,0.0)	{\includegraphics[scale=1]{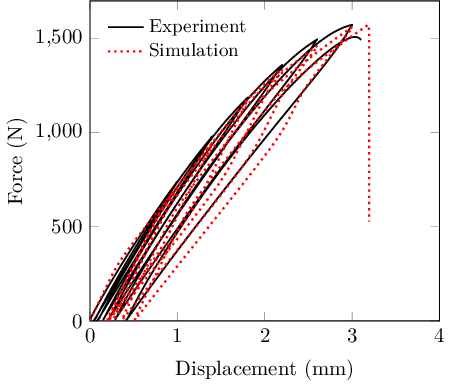}};
		\end{tikzpicture}
		\caption{}
		\label{fig:force_disp_np1_w0}
	\end{subfigure}%
	\begin{subfigure}[b]{0.5\textwidth}
		\centering
		\begin{tikzpicture}
			\node[inner sep=0pt]  at (0.0,0.0)	{\includegraphics[scale=1]{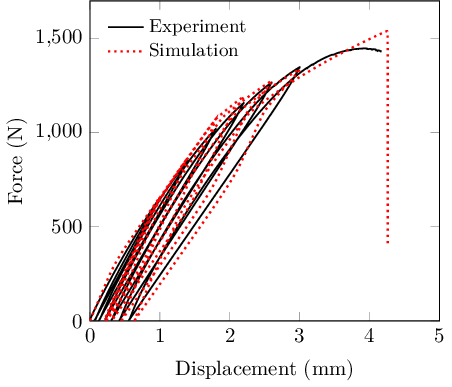}};			
		\end{tikzpicture}
		\caption{}
		\label{fig:force_disp_np1_w1}
	\end{subfigure}%
	\caption{Effect of moisture on the force–displacement response of BNP(10~\%wt)/epoxy at room temperature and the deformation rate of 1~mm/min: (a) dry sample, and (b) saturated sample.}
	\label{fig:db_np10}
\end{figure}

Next, dogbone tensile tests are conducted to evaluate the capability of the PFM model in predicting the fracture behavior of BNP/epoxy nanocomposites with 10~\%wt of BNPs. The simulations are performed using the calibrated material parameters listed in Tables~\ref{tab:param}, and the effect of nanoparticle contents is taken into account using the amplification factor presented in Eq.~\ref{eq:X}. Figs.~\ref{fig:force_disp_np1_w0}~and~\ref{fig:force_disp_np1_w1} show the effect of BNP weight fraction on the force-displacement response at the deformation rate of $\dot{u}=1$~mm/min and $\Theta=296$~K. Both experimental data and numerical predictions are presented in the figures. In these curves, good agreements are observed between the numerical results and experimental data, evidencing the practicability of the PFM at different BNP weight fractions.

\begin{figure}[!h]
	\centering
		\centering
		\begin{tikzpicture}
			\node[inner sep=0pt]  at (0.0,0.0)	{\includegraphics[scale=1]{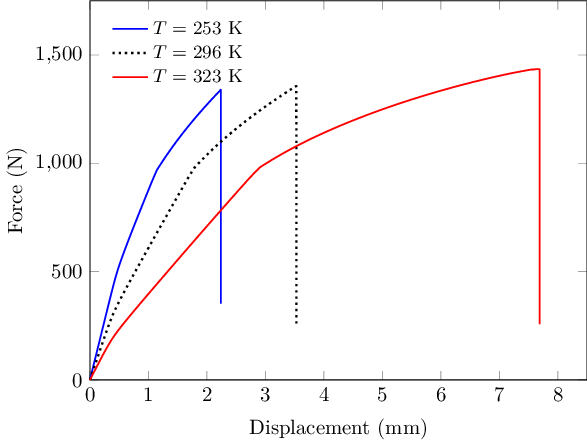}};
		\end{tikzpicture}
		\caption{Effect of temperature on the force–displacement curve in dogbone simulation tests of saturated BNP(10~\%wt)/epoxy samples at the deformation rate of 1~mm/min.}
		\label{fig:force_disp_np1_w01_temp}
\end{figure}

We then explore how temperature affects the fracture behavior of BNP/epoxy nanocomposites. The results of our dogbone simulation tests, presented in Fig.~\ref{fig:force_disp_np1_w01_temp}, show that specimens with 10~\%wt of BNPs fractures at different displacement ranges depending on the temperature: around 2.2~mm at 253~K, around 5~mm at 296~K, and around 7.6~mm at 323~K. These observations can interpreted in two ways. First, as temperature increases, the shear and bulk modulus associated with the equilibrium, non-equilibrium, and volumetric responses decrease (see Eqs.~\ref{eq:mu_0} and \ref{eq:mu_e0}), resulting in a less stiff material. Second, the deformation of the viscoelastic composite is temperature-dependent, meaning that an increase in temperature leads to a greater displacement at the fracture initiation due to a rise in the viscous flow. 

It is important to note that the simulations assume the effect of temperature on the energy release rate to be negligible. While this assumption is acceptable based on experimental observations, further numerical-experimental validations are recommended to identify the variation of energy release rate with temperature accurately.

\begin{figure}[!h]
	\centering
	\begin{subfigure}[b]{0.35\textwidth}
			\centering			
			\begin{tikzpicture}
				\node[inner sep=0pt]  at (0.0,0.0)	{\includegraphics[scale=1]{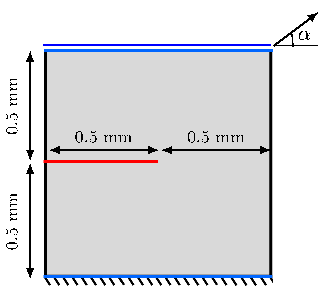}};
			\end{tikzpicture}
			\caption{ }
			\label{fig:sn_a}
		\end{subfigure}%
	\begin{subfigure}[b]{0.3\textwidth}
			\centering			
			\begin{tikzpicture}
				\node[inner sep=0pt]  at (0.0,0.0)	{\includegraphics[scale=1]{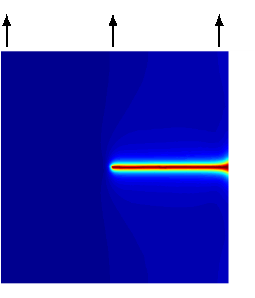}};
			\end{tikzpicture}
			\caption{ }
			\label{fig:sn_b}
		\end{subfigure}%
	\begin{subfigure}[b]{0.3\textwidth}
			\centering			
			\begin{tikzpicture}
				\node[inner sep=0pt]  at (0.0,0.0)	{\includegraphics[scale=1]{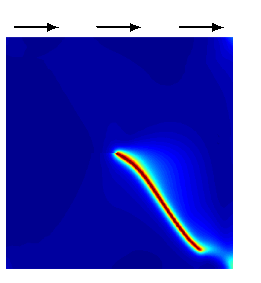}};
			\end{tikzpicture}
			\caption{ }
			\label{fig:sn_c}
		\end{subfigure}%
	\caption{(a) Geometry and boundary conditions of single edge notched specimen, (b) fracture pattern for unidirectional tension ($\alpha=90^{\circ}$), and (c) fracture pattern for pure shear deformation ($\alpha=0^{\circ}$). The deformation rate is 1~mm/min.}
	\label{fig:sn_contour}	
\end{figure}

The proposed model's ability to predict fracture patterns is evaluated by conducting single-edge notched tensile and shear tests on pure epoxy samples. The geometry and boundary conditions of the tests are shown in Figure~\ref{fig:sn_a}. A horizontal notch is placed at the center of the specimen's left outer surface. The bottom side of the specimen is fixed, while the top side is moved. Both tensile and shear loads are applied at a deformation rate of $\dot{u}=1$~mm/min with constant displacement increments of $10^{-6}$~mm. The simulations are performed at 296~K under plane strain conditions using the material parameters listed in Table~\ref{tab:param}, while the scale parameter is taken to be 0.015~mm. Meshes are refined in areas where cracks are expected to propagate. 

For the tensile test, a discretization with 12509 elements and an effective element size of 0.003~mm is generated in the central strip of the specimen. To reach the same element size at the critical zone in the shear test, 21045 elements with refined meshes in the lower right diagonal strip of the specimen are used. The fracture patterns for the two cases are shown in Figures~\ref{fig:sn_b} and \ref{fig:sn_c}. It can be observed that the crack path is horizontal for the tensile case, while there is a curved crack path for the pure shear case. From the shear test in Figure~\ref{fig:sn_c}, it can be found that the free energy decomposition presented in Eq.~\eqref{eq:free_energy_decomp} prevents cracking in compression. The crack patterns obtained are consistent with those reported in the literature~\cite{miehe2010phase}.

\section{Summary and conclusions}
\label{sec:summary}

A finite deformation phase-field fracture model has been developed to study the cyclic viscoelastic-viscoplastic fracture behavior of BNPs reinforced epoxy nanocomposites under hygrothermal conditions. For this, the PFM with a modified crack driving force has been coupled to a viscoelastic-viscoplastic constitutive law. Within the derived formulation, a free energy has additively been decomposed into an equilibrium, a non-equilibrium, and a volumetric part with positive/negative components, capturing the effect of nanoparticles, moisture and temperature on the nonlinear material behavior of the nanocomposites. Also, an amplification factora modified version of the Kitagawa model have been adopted to capture the role played by nanoparticles, moisture and temperature on the fracture behavior. 

The proposed PFM has been implemented in the FE analysis of dogbone tensile tests to demonstrate its applicability. Numerical simulations were conducted to investigate the impact of nanoparticle content on the force-displacement response of dry and saturated BNP/epoxy samples subjected to cyclic loading-unloading. The results indicated that the force-displacement responses obtained from the numerical simulations were consistent with those of experimental tests at various nanoparticle and moisture contents. The comparison between numerical and experimental results confirms the capacity of the PFM in predicting the development of damage and viscoplasticity in BNP/epoxy nanocomposites at different nanoparticle and moisture contents.

To evaluate the proposed PFM's potential for broader applications, it would be worthwhile to compare the numerical predictions at different temperatures and deformation rates with experimental data in the future. Also, the physical and chemical interactions between nanoparticles and epoxy matrices undergo significant changes under finite deformation at the micro- and sub-micro-scale. These changes would impact other material properties, such as viscosity and energy release rate. However, due to the complex interactions at the interphase of nanoparticle/epoxy, the mechanisms leading to these changes are not clearly understood. In the proposed model, the changes are considered by taking the volume fraction of nanoparticles and moisture contents. To gain a deeper understanding of the microstructure's effect on the macroscopic properties, the PFM can be informed by molecular simulations~\cite{arash2015tensile,arash2016coarse}. It would allow the analysis of the polymer nanocomposites at the molecular scale.

Furthermore, the aggregation of nanoparticles in the epoxy matrix causes insufficient dispersal~\cite{khorasani2019effect}, which results in the degradation of material properties due to relatively inferior interactions between the nanoparticles and the matrix. A study on the effect of surface modification of BNPs is suggested for future research to suppress the aggregation and enhance the interfacial properties. The reinforcement of the epoxy with surface-modified BNPs can improve the fracture properties of the nanocomposites.

\section*{Acknowledgments}

This work originates from two research projects: (1) ”Hybrid laminates and nanoparticle-reinforced materials for improved rotor blade structures” (”LENAH - Lebensdauererhöhung und Leichtbauoptimierung durch nanomodifizierte und hybride Werkstoffsysteme im Rotorblatt”), funded by the Federal Ministry of Education and Research of Germany, and (2) ”Challenges of industrial application of nanomodified and hybrid material systems in lightweight rotor blade construction” (”HANNAH - Herausforderungen der industriellen Anwendung von nanomodifizierten und hybriden Werkstoffsystemen im Rotorblattleichtbau”), funded by the Federal Ministry for Economic Affairs and Climate Action. The authors wish to express their gratitude for the financial support.

%The authors acknowledge the support of the LUIS scientific computing cluster, which is funded by Leibniz Universität Hannover, the Lower Saxony Ministry of Science and Culture (MWK) and the German Research Council (DFG).
%%
%%
%%\section*{Data Availability}
%%
%%The raw/processed data required to reproduce these findings cannot be shared at this time as the data also forms part of an ongoing study.

\bibliography{mybibfile}

\end{document}